\documentclass[letterpaper]{article} 
\usepackage{graphicx} 
\usepackage{aaai24}  
\usepackage{times}  
\usepackage{helvet}  
\usepackage{courier}  
\usepackage{multirow}
\usepackage{xcolor}
\usepackage{booktabs}
\usepackage{adjustbox}
\usepackage{xcolor,colortbl}
\definecolor{bubbles}{rgb}{0.91, 1.0, 1.0}

\definecolor{Gray}{gray}{0.85}
\definecolor{LightCyan}{rgb}{0.88,1,1}
\definecolor{lightthulianpink}{rgb}{0.9, 0.56, 0.67}
\definecolor{mypink3}{cmyk}{0, 0.7808, 0.4429, 0.1412}
\newcolumntype{a}{>{\columncolor{Gray}}c}
\newcolumntype{b}{>{\columncolor{white}}c}

\definecolor{blizzardblue}{rgb}{0.67, 0.9, 0.93}
\usepackage[hyphens]{url}  
\usepackage{natbib}  
\usepackage{caption} 
\usepackage{algorithm}
\usepackage{algorithmic}

\usepackage{bibentry}

\usepackage{newfloat}
\usepackage{listings}
\DeclareCaptionStyle{ruled}{labelfont=normalfont,labelsep=colon,strut=off} 
\floatstyle{ruled}
\newfloat{listing}{tb}{lst}{}
\floatname{listing}{Listing}
\title{Enhancing Recommender Systems with Large Language Model Reasoning Graphs}

\author{
    Yan Wang\equalcontrib\textsuperscript{\rm 1},
    Zhixuan Chu\equalcontrib\textsuperscript{\rm 1}\thanks{Corresponding author.},
    Xin Ouyang\textsuperscript{\rm 1},
    Simeng Wang\textsuperscript{\rm 1},
    Hongyan Hao\textsuperscript{\rm 1},
    Yue Shen\textsuperscript{\rm 1},
    Jinjie Gu\textsuperscript{\rm 1},
    Siqiao Xue\textsuperscript{\rm 1},
    James Y Zhang\textsuperscript{\rm 1},
    Qing Cui\textsuperscript{\rm 1},
    Longfei Li\textsuperscript{\rm 1},
    Jun Zhou\textsuperscript{\rm 1},
    Sheng Li\textsuperscript{\rm 2}
}
\affiliations{
    \textsuperscript{\rm 1}Ant Group\\
    \textsuperscript{\rm 2}University of Virginia\\
\ \{luli.wy, chuzhixuan.czx, xin.oyx, simeng.wsm, hongyanhao.hhy, zhanying, jinjie.gujj, siqiao.xsq, james.z, cuiqing.cq, longyao.llf, jun.zhoujun\}@antgroup.com, shengli@virginia.edu 

}

\begin{document}

\maketitle

\begin{abstract}
Recommendation systems aim to provide users with relevant suggestions, but often lack interpretability and fail to capture higher-level semantic relationships between user behaviors and profiles. In this paper, we propose a novel approach that leverages large language models (LLMs) to construct personalized reasoning graphs. These graphs link a user's profile and behavioral sequences through causal and logical inferences, representing the user's interests in an interpretable way. Our approach, LLM reasoning graphs (LLMRG), has four components: chained graph reasoning, divergent extension, self-verification and scoring, and knowledge base self-improvement. The resulting reasoning graph is encoded using graph neural networks, which serves as additional input to improve conventional recommender systems, without requiring extra user or item information. Our approach demonstrates how LLMs can enable more logical and interpretable recommender systems through personalized reasoning graphs. LLMRG allows recommendations to benefit from both engineered recommendation systems and LLM-derived reasoning graphs. We demonstrate the effectiveness of LLMRG on benchmarks and real-world scenarios in enhancing base recommendation models. 
\end{abstract}

\section{Introduction}
Recommendation systems are now prevalent across the internet, smartly surfacing personalized contents and products to users based on their individual profiles and historical behavioral data \cite{geng2022recommendation,hui2022personalized,chu2022hierarchical,li2021survey,jiang2022learning}. However, the vast majority of recommendation systems rely solely on conventional machine learning techniques, which can only identify patterns and relationships within sequences of interactions without actually comprehending the true meaning or semantics behind the items themselves. Devoid of any logical or causal reasoning capacities, these recommender systems struggle to effectively capture the full spectrum of conceptual relationships and connections spanning a user's diverse interests and behavioral patterns over time.

In addition, recent work \cite{wang2019knowledge,chen2021temporal,wu2019session,wang2020disentangled} has sought to enhance recommendations by incorporating graph-structured information, which provides valuable contextual data beyond standard user-item interactions. However, even these more advanced knowledge graph-based recommendation systems still lack the ability to perform complex reasoning or inference - simply overlaying factual relationships is not enough to enable a system to deeply understand users' interests and generate truly insightful recommendations.

In parallel, tremendous progress in large language models (LLMs) \cite{radford2018improving,radford2019language,brown2020language,ouyang2022training} like GPT-3, GPT-4, Claude, and others has demonstrated powerful new capacities for reasoning, inference, and logic without the need for explicit training on such tasks. These models exhibit remarkable aptitudes for causal, logical, and analogical reasoning, illuminating new opportunities to leverage their strengths to develop superior knowledge representations that can capture nuanced semantic relationships between users' interests \cite{sheu2021knowledge}. By leveraging LLMs to reason behavioral sequences and comprehend user interests at a deeper conceptual level, there is immense potential to revolutionize next-generation recommendation systems.

Therefore, we propose using an LLM to construct personalized reasoning graphs for recommendation systems. The LLM inputs a user's profile and behavioral sequences and outputs a graphical representation linking concepts through chained causal and logical reasoning. This results in an expansive graph that encodes higher-level semantic relationships between the user's interests and behaviors. We then apply SR-GNN \cite{wu2019session} to learn a dense feature representation that summarizes the graph's structure and semantics. This graph embedding is provided as additional input to the conventional recommendation models, such as BERT4Rec \cite{sun2019bert4rec}, FDSA \cite{zhang2019feature}, CL4SRec \cite{xie2022contrastive}, and DuoRec \cite{qiu2022contrastive}. Our approach allows recommendations to consider conceptual relationships derived through reasoning while still benefiting from the recommendation abilities of traditional models. Moreover, the graph provides interpretability by surfacing the explicit reasoning behind recommendations. 

We designed four interlocking modules, powered by LLMs, to construct personalized reasoning graphs that model each user's interests: 1) a chained graph reasoning module that conducts chained causal and logical reasoning, 2) a divergent extension module that expands the graph by associating and reasoning about the user's interests, 3) a self-verification and scoring module that validates the reasoning procedure through abductive reasoning and scoring, and 4) a knowledge base self-improving module that caches validated reasoning chains for later reuse. Together, these four modules construct Large Language Model Reasoning Graphs (LLMRG) paradigm, which employs a prompt-based framework leveraging LLMs to imaginatively generate plausible new reasoning chains, given their behavioral history and features. Besides, LLMRG can perform imaginary continuations of each reasoning chain to predict the next items the user is likely to engage with. This divergent thinking allows us to go beyond reactive recommendations based on consumed content to proactively recommend new items tailored to modeling the user's intention.

Experiments demonstrate our model's ability to improve recommendation performance without requiring additional user or item data. This work illustrates how large language models can enable logical and interpretable recommender systems. In summary, our model uniquely achieves more reasoned recommendations by leveraging LLMs to construct personalized reasoning graphs that capture causal and logical inferences about users' profiles and behaviors.

\section{Background}

\subsection{Graph-based recommendation system}

Recent work explores graph-based methods that can incorporate additional relationship information into recommendation systems. For example, knowledge graphs have emerged as a powerful way to represent relationships between entities to capture complex entity interactions \cite{wang2019knowledge,chen2021temporal,chu2021graph,sheu2021knowledge,chu2024llm}. Beyond predefined knowledge graphs, some methods \cite{wang2019knowledge} learn to construct an informative graph from user-item interactions. While knowledge graphs provide external information, graph learning methods \cite{wu2019session} can extract latent structures. Combining the two concepts, \cite{wang2020disentangled} jointly leverage a knowledge graph and interaction graph. In summary, graph-based methods allow recommendation models to encode richer connectivity patterns. However, there are some potential disadvantages of graph-based recommendation systems compared to reasoning graph construction by LLMs: (1) Knowledge graphs require extensive human expertise to build and maintain relationships, whereas LLMs can automatically extract relational knowledge from large text corpora; (2) Predefined knowledge graphs may have coverage gaps for certain entities or domains. LLMs can learn to reason about any entity mentioned in the text; (3) Graph learning methods that construct graphs from interactions are limited to observable user-item connections. LLMs can infer more abstract and latent relationships through reasoning; (4) Knowledge graphs and graphs are static after construction. LLMs can continue to expand their knowledge and reasoning capabilities as they are trained on more data.

\subsection{Reasoning of LLM}
Recent advances in large language models (LLMs) like GPT-3 and PaLM have enabled strong capabilities in logical and causal reasoning \cite{zhong2023chatabl, shi2023language,yoneda2023llm,yao2021survey,chu2023causal,gu2023robust}. This progress stems from three key strengths. First, natural language understanding allows LLMs to parse meaning and relationships from text \cite{devlin2018bert,brown2020language}. Models can identify entities, actions, and causal chains through techniques like self-attention and contextual embeddings \cite{zhao2022tiny,xie2022nc}. For example, BERT uses masked language modeling to learn bidirectional representations that incorporate context. Second, LLMs have accumulated vast commonsense knowledge about how the world works \cite{shin2021constrained,chowdhery2022palm,zhao2023logic,tsai2023game}. GPT-3 was trained on over a trillion words from the internet, absorbing implicit knowledge about physics, psychology, and reasoning. Models like PaLM were further trained with constrained tuning to better incorporate common sense. This enables filling in missing premises and making deductions. Third, transformer architectures impart combinatorial generalization and symbolic reasoning abilities \cite{wei2022chain}. Self-attention layers allow LLMs to chain ideas, follow arguments step-by-step, and make coherent deductions. For example, Chain of Thought prompts GPT-3 to explain its reasoning for robustness. Together, these strengths of understanding language, leveraging knowledge, and combinatorial reasoning empower LLMs to parse scenarios, tap relevant knowledge, and reason through implications and causes.

\section{LLMRG}
\subsection{Problem Statement}

\begin{figure*}[t]
    \centering
    \includegraphics[width=1.8\columnwidth]{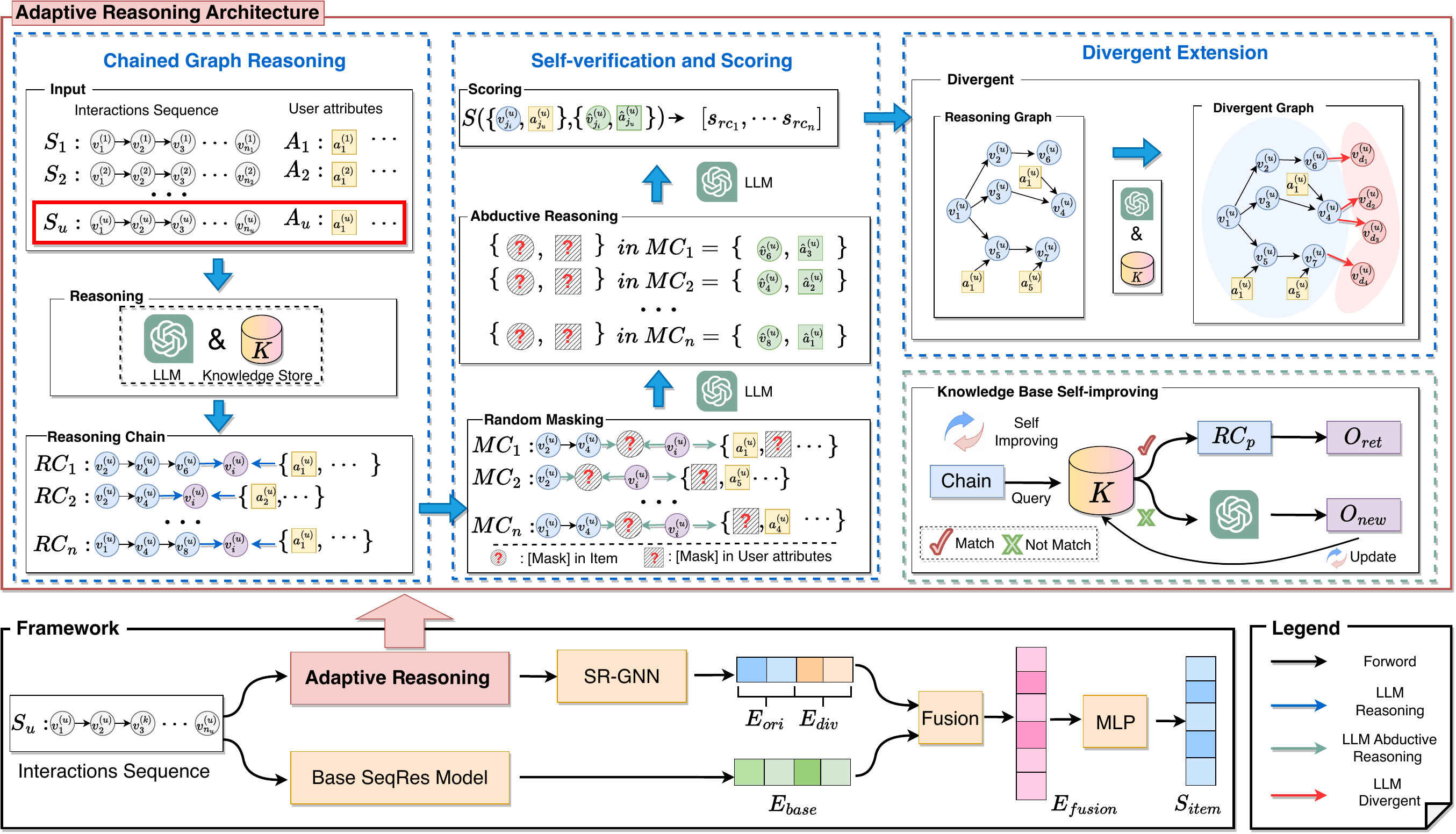}
    
    \caption{LLMRG framework has two main components, i.e., an adaptive reasoning module with self-verification and a base sequential recommendation model. Our model concatenates the embeddings from the adaptive reasoning module ($E_{ori}$  and $E_{div}$) and the base model ($E_{base}$) to obtain $E_{fusion}$. This fused embedding is used to predict the next item for the user. The key advantage of our approach is that the adaptive reasoning module can construct personalized reasoning graphs, going beyond the sequential modeling of user interests. The self-verification and scoring help improve the reasoning process. Fusing this with a standard recommendation model allows for combining complementary strengths without accessing extra information.}
    
    \label{fig:framework}
\end{figure*}

Recommender systems aim to predict users' interests based on their historical interactions. Sequential recommendation approaches this as a sequential modeling problem - viewing the user's history of interactions as an ordered sequence and attempting to model the user's dynamically evolving interests. Formally, let $\mathcal{U}{=}\{u_1,u_2,\dots,u_{\vert\mathcal{U}\vert}\}$ denote the set of users, $\mathcal{V}{=}\{v_1,v_2,\dots,v_{|\mathcal{V}|}\}$ be the set of items, and list $\mathcal{S}_u{=}[v_1^{(u)},\dots,v_t^{(u)},\dots,v_{n_u}^{(u)}]$ denote the sequence of interactions for user $u \in \mathcal{U}$ in chronological order, where $v_t^{(u)} \in \mathcal{V}$ is the item interacted with at time step $t$ and $n_u$ is the length of the sequence. We use relative time indices instead of absolute timestamps. In addition, let $\mathcal{A}_u{=}[a_1^{(u)},\dots,a_i^{(u)},\dots,a_{n_a}^{(u)}]$ represent user attributes for modeling personalization, where $n_a$ is the number of attributes. Given a user's interaction history $\mathcal{S}_u$, the sequential recommendation task is to predict the item user $u$ will interact with at the next time step $n_{u} + 1$. This can be formalized as modeling the probability distribution over all possible items for user $u$ at time step $n_{u} {+} 1$:
\begin{equation}
    p\bigl(v_{n_u+1}^{(u)}=v|\ \mathcal{S}_u, \mathcal{A}_u\bigr).
\end{equation}
In this work, we propose constructing Large Language Model Reasoning Graphs (LLMRG), a new paradigm that utilizes LLMs to improve recommendation system performance. We first use a large language model (LLM) to construct personalized reasoning graphs based on $\mathcal{S}_u$ and $\mathcal{A}_u$, which reason a user's profile and behavioral sequences through causal and logical inferences. The graph provides an interpretable model of a user's interests and embeds rich semantic relationships. We propose an adaptive reasoning architecture with self-verification based on the capabilities of LLMs, which includes four components: 1) chained graph reasoning, 2) divergent extension, 3) self-verification and scoring, and 4) self-improved knowledge base. By encoding the resulting conceptual reasoning graph using graph neural networks, it can be provided as an additional input into conventional recommender systems. This allows recommendations to benefit from both engineered recommendation algorithms and the explanatory knowledge derived from the LLM graph reasoning process. In this section, we will detail this whole framework, and the detailed prompt examples of each module are illustrated in the Appendix.

\subsection{Adaptive Reasoning Architecture}

\paragraph{Chained Graph Reasoning.}
Along with the user behavioral sequences $\mathcal{S}_u$, for each item, we construct reasoning chains ${RC}_n$ that link it to existing chains if there are logical connections or start entirely new chains rooted in the item itself if there are no applicable links to existing reasoning chains. Relevant user attributes $\mathcal{A}_u$ are incorporated where possible to further customize the reasoning chains for recommending items. This iterative reasoning chain construction process is carried out progressively along the user's behavioral sequence up until the last item. Specifically, we employ a prompt-based framework leveraging large language models to imaginatively generate plausible new reasoning chains that could logically motivate the user to engage with the next known item in their sequences. The prompt takes as input the known next item, existing reasoning chains constructed thus far, and available user attributes. It outputs a comprehensive set of possible new reasoning chains explaining why the user might want to take the next item. These dynamically generated new chains are integrated into the evolving logical reasoning graph to enable the modeling of increasingly complex interdependent motivations and interests underlying the user's evolving behavioral trajectory.

\paragraph{Divergent Extension.}

Besides the observed behavioral sequences, we aim to conduct divergent thinking according to the established reasoning graph. We propose a new divergent extension module that performs imaginary continuations of each reasoning chain to predict the next items the user is likely to engage with. Specifically, for each reasoning chain digging into the user's motivations and thinking process, the divergent extension module employs an imagination engine to divergently extend the chain beyond the last known item. This involves using the language model to sample plausible continuations of the reasoning trajectory that predict what other related items the user might be interested in next. For example, if the chain represents an interest in sci-fi movies with complex philosophies, the extension could generate new sequences predicting more cerebral sci-fi films with similar themes and tones that the user might enjoy. Critically, the imagination engine outputs multiple diverse possible extending items per reasoning chain, capturing the user's multifaceted interests. These imaginary new items represent predictions of movies the user is likely to watch soon. We aggregate the predicted new items from all the extended reasoning chains to form a comprehensive set of personalized recommendations tailored to the user's preferences. It is worth noting that the generated new item recommendations may not exist in the original item list for our recommendation task. Therefore, we need to use another small language model to calculate the similarity between the generated items and the original list in order to retrieve the most relevant item recommendations. The divergent thinking allows us to go beyond reactive recommendations based on consumed content to proactively recommend new items tailored to modeling the user's motivations. In this procedure, a prompt-based framework based on LLM is still employed. It is worth noting that rather than just predicting the single next movie, our divergent extension module enables generating multiple future trajectories per reasoning chain. This allows for properly capturing the user's diverse interests and possibilities they may take next.

\paragraph{Self-verification and Scoring.}
The self-verification module utilizes the abductive reasoning capability \cite{xu2023large} of LLM to check the plausibility and coherence of the dynamically generated reasoning chains from the chained graph reasoning and divergent extension modules. Before adding a new reasoning chain to the graph, the module masks the key items or engaged user attributes that the chain is meant to logically link to. It then prompts the large language model to fill in the [Mask] in the masked chains $MC_n^{(u)}$ with the most reasonable prediction. If the predicted item or attribute matches what was originally masked, this provides evidence that the reasoning chain logically flows and is consistent with the user's behavioral history and attributes. The higher the match score, the more robust the reasoning graph is as a whole. On the other hand, a low match score indicates potential flaws in the coherence or plausibility of some reasoning chains. The system can then selectively filter out or recalibrate the problematic chains before integrating them into the graph. Therefore, we set a threshold score for this self-verification to judge the rationality of reasoning. This improves the overall soundness of the dynamically constructed reasoning chains for the chained graph reasoning and divergent extension modules, ensuring reliable reasoning for recommendations aligned with the user's interests. Specifically, this module mainly involves three steps, i.e., random masking, abductive reasoning, and scoring, which are exemplified in Figure \ref{fig:framework}.

\paragraph{Knowledge Base Self-improving.} In our system's chained graph reasoning, divergent extension, and self-verification modules, we make extensive use of a language model to conduct inference and reasoning. This repeated language model invocation incurs significant computational costs. However, we observed that many knowledge elements and reasoning procedures are applied repeatedly across queries. To avoid redundant work, we introduce a knowledge base that caches validated reasoning chains for later reuse. By reusing previous reasoning results rather than re-computing them, we substantially reduce language model usage. We employ a self-improving approach to maintain knowledge base quality over time. Using the scores from our self-verification and scoring module, which assess reasoning chain validity, we retain only high-quality chains in the knowledge base. Low-scoring chains are discarded to filter out low-quality or erroneous inferences. Before conducting new reasoning, we first check whether the knowledge base already contains a relevant chain. If so, we retrieve and leverage that pre-computed chain instead of invoking the language model. This knowledge base of cached, high-quality reasoning chains significantly reduces computational requirements. Our experiments demonstrate it can cut language model usage by about $30\%$ compared to inferences from scratch after $3000$ times of reasoning and verification steps in Figure \ref{fig:exp_self_improving}.

\begin{figure*}[t]
    \centering
    \includegraphics[width=1.9\columnwidth]{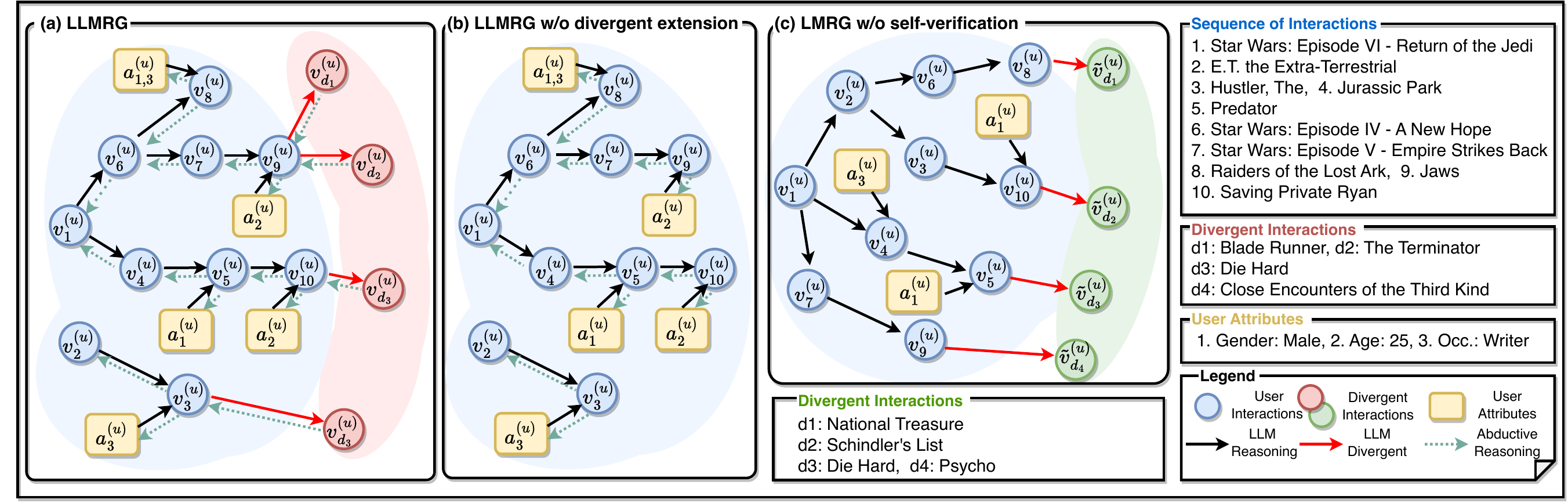}
    
    \caption{The real case studies (ML-1M) on our (a) LLMRG and ablation models, i.e., (b) LLMRG w/o divergent extension and (c) LLMRG w/o self-verification. The black arrow represents the reasoning procedure. The red arrow is the divergent extension. The green dashed arrow refers to the abductive reasoning in the self-verification module.}
    
    \label{fig:real_case_example}
\end{figure*}

\subsection{LLMRG Framework}
Sequential recommendation approaches typically view the user's history of interactions as an ordered sequence and attempt to model the user's dynamically evolving interests. In this work, we propose to use an LLM to construct personalized reasoning graphs for recommendation systems. Therefore, as shown in Figure \ref{fig:framework}, our proposed LLMRG has two components, i.e., an adaptive reasoning module with self-verification and a base sequential recommendation model.

The adaptive reasoning module takes the user's interaction sequence  $\mathcal{S}_u{=}[v_1^{(u)},\dots,v_t^{(u)},\dots,v_{n_u}^{(u)}]$ and attributes $\mathcal{A}_u{=}[a_1^{(u)},\dots,a_i^{(u)},\dots,a_{n_a}^{(u)}]$ as input. This input goes through chained graph reasoning, self-verification and scoring, and divergent extension repeatedly to construct a reasoning graph and a divergent graph. The adaptive reasoning module is expressed as a mapping $\phi: \{\mathcal{S}_u, \mathcal{A}_u  \} \rightarrow \{G_{rea}, G_{div}$\}, where $G_{rea}$ and $G_{div}$ represent the reasoning graph and divergent graph, respectively. We utilize SR-GNN \cite{wu2019session} to automatically extract embeddings from the graphs, which can support different connection matrices and directed graphs in the recommendation task. This process produces two embeddings $E_{ori}$  for the reasoning graph and $E_{div}$ for the divergent graph by $g_1: G_{rea} \rightarrow E_{ori}$ and $g_2: G_{div} \rightarrow E_{div}$. In parallel, the base sequential recommendation model directly processes the input to produce an embedding $E_{base}$. Finally, we concatenate the embeddings from the adaptive reasoning module ($E_{ori}$  and $E_{div}$) and the base model ($E_{base}$) to obtain $E_{fusion}$. This fused embedding is used to predict the next item for the user by $\psi: E_{fusion} \rightarrow v_{n_u+1}^{(u)}$.

The key advantages of our approach are that the adaptive reasoning module can construct personalized reasoning graphs, and the divergent extension employs divergent thinking to go beyond reactive recommendations to proactively recommend new items following the evolving behavioral trajectory. The self-verification and scoring also help improve the reasoning process. Fusing this with a standard sequential recommendation model allows for combining complementary strengths without accessing extra information.

\begin{table}

    \centering
    \caption{Statistics of the datasets after preprocessing.}
    
    \resizebox{0.65\columnwidth}{!}{
        \begin{tabular}{l| ccc}
        \toprule
        Specs. & Beauty & Clothing  & ML-1M\\
        \midrule
       \# Users & 22,363 & 39,387  &  6,041 \\
       \# Items & 12,101 & 23,033  & 3,417 \\
       \# Avg.Length & 8.9 & 7.1  & 165.5\\
       \# Actions & 198,502 & 278,677  & 999,611\\
       Sparsity & 99.93\% & 99.97\%  & 95.16\%\\
       \bottomrule
    \end{tabular}}
    
    \label{Datasets}
\end{table}

\section{Experiments}

\subsection{Experiment Settings}

\begin{table*}[t!]
    \centering
    
    \caption{Performance comparison on three benchmark datasets, i.e., ML-1M, Amazon Beauty, and Amazon Clothing. We set the original models as baselines to compare with our proposed LLMRG model based on GPT3.5 or GPT4. The shaded area indicates the improved performance of our LLMRG model over the baselines across all three datasets. Higher is better.}
    
    \resizebox{1.9\columnwidth}{!}{
        \begin{tabular}{l l|baa |baa |baa |baa}
        \toprule
        \multirow{2.5}{*}{Dataset} & \multirow{2.5}{*}{Metric} &
        \multicolumn{3}{c}{FDSA} & \multicolumn{3}{c}{BERT4Rec} & 
        \multicolumn{3}{c}{CL4SRec} & \multicolumn{3}{c}{DuoRec} \\
        
          \cmidrule(lr){3-5} \cmidrule(lr){6-8} \cmidrule(lr){9-11} \cmidrule(lr){12-14}
        
          &  & Original  & GPT3.5 & GPT4 
         & Original  & GPT3.5 & GPT4
         & Original  & GPT3.5 & GPT4
         & Original  & GPT3.5 & GPT4\\
       
        \midrule
         \multirow{4}{4em}{ML-1M} & HR@5 & 0.0909  &  \textcolor{red}{+ 20.70\%} & \textcolor{red}{+ 25.79\%} & 0.1124 & \textcolor{red}{+ 26.67\%} & \textcolor{red}{+ 32.56\%} & 0.1141 & \textcolor{red}{+ 19.98\%} & \textcolor{red}{+ 21.02\%} & 0.2011 & \textcolor{red}{+ 12.87\%}&  \textcolor{red}{+ 14.76\%}\\
         & HR@10 & 0.1631 & \textcolor{red}{+ 17.93 \%} & \textcolor{red}{+ 22.87\%} & 0.1910 & \textcolor{red}{+ 13.52 \%} & \textcolor{red}{+ 16.49 \%} & 0.1866 & \textcolor{red}{+ 17.30 \%} & \textcolor{red}{+ 19.31 \%} & 0.2837 & \textcolor{red}{+ 14.10 \%} & \textcolor{red}{+ 15.53 \%} \\
        & NDCG@5 & 0.0599 & \textcolor{red}{+ 21.33\%} & \textcolor{red}{+ 30.27 \%} & 0.0713 & \textcolor{red}{+ 25.74 \%} & \textcolor{red}{+ 32.82 \%} & 0.0721 & \textcolor{red}{+ 14.97 \%} & \textcolor{red}{+ 16.78 \%} & 0.1265 & \textcolor{red}{+ 23.55 \%} & \textcolor{red}{+ 26.01 \%} \\
        & NDCG@10 & 0.0878 & \textcolor{red}{+ 21.78\%} & \textcolor{red}{+ 28.25\%} & 0.0980 & \textcolor{red}{+ 23.34\%} & \textcolor{red}{+ 28.06\%} & 0.1013 & \textcolor{red}{+ 17.67\%} & \textcolor{red}{+ 20.42\%} & 0.1663 & \textcolor{red}{+ 12.86\%} & \textcolor{red}{+ 13.77\%} \\
         \midrule
        & HR@5 & 0.0237 & \textcolor{red}{+ 13.89 \%} & \textcolor{red}{+ 17.53 \%} & 0.0201 & \textcolor{red}{+ 19.17 \%} & \textcolor{red}{+ 23.22 \%} & 0.0398 & \textcolor{red}{+ 11.15 \%} & \textcolor{red}{+  14.15 \%} & 0.0552 & \textcolor{red}{+ 9.31 \%} & \textcolor{red}{+ 11.93 \%} \\
        Amazon & HR@10 & 0.0418  & \textcolor{red}{+ 15.02 \%} & \textcolor{red}{+ 17.78 \%} & 0.0413 & \textcolor{red}{+ 17.79 \%} & \textcolor{red}{+ 22.14 \%} & 0.0664 & \textcolor{red}{+ 10.22 \%} & \textcolor{red}{+ 11.32 \%} & 0.0839 & \textcolor{red}{+ 5.14 \%} & \textcolor{red}{+ 6.61 \%} \\
        Beauty & NDCG@5 & 0.0195 & \textcolor{red}{+ 16.20 \%} & \textcolor{red}{+ 18.64 \%} & 0.0192 & \textcolor{red}{+ 14.21 \%} & \textcolor{red}{+ 17.63 \%} & 0.0221 & \textcolor{red}{+ 8.45 \%} & \textcolor{red}{+ 10.18 \%} & 0.0350 & \textcolor{red}{+ 7.42 \%} & \textcolor{red}{+ 9.24 \%} \\
        & NDCG@10 & 0.0275 & \textcolor{red}{+ 14.78 \%} & \textcolor{red}{+ 17.64 \%} & 0.0263 & \textcolor{red}{+ 11.53 \%} & \textcolor{red}{+ 14.76 \%} & 0.0322 & \textcolor{red}{+ 8.17 \%} & \textcolor{red}{+ 9.68 \%} & 0.0447 & \textcolor{red}{+ 6.67 \%} & \textcolor{red}{+ 7.95 \%} \\
        \midrule
        & HR@5 &  0.0119 & \textcolor{red}{+ 20.67 \%} & \textcolor{red}{+ 23.92 \%} & 0.0128 & \textcolor{red}{+ 16.09 \%} & \textcolor{red}{+ 19.10 \%} & 0.0166 & \textcolor{red}{+ 7.90 \%} & \textcolor{red}{+ 10.92 \%} & 0.0190 & \textcolor{red}{+ 9.98 \%} & \textcolor{red}{+ 11.40 \%} \\
       Amazon  & HR@10 & 0.0197 & \textcolor{red}{+ 14.45 \%} & \textcolor{red}{+ 17.88 \%} & 0.0202 & \textcolor{red}{+ 10.52 \%} & \textcolor{red}{+ 13.72 \%} & 0.0273 & \textcolor{red}{+ 11.21 \%} & \textcolor{red}{+ 14.99 \%} & 0.0311 & \textcolor{red}{+ 7.65 \%}  &  \textcolor{red}{+ 9.48 \%}  \\
    Clothing & NDCG@5 & 0.0073 & \textcolor{red}{+ 8.16 \%} & \textcolor{red}{+ 10.86 \%} & 0.0081 & \textcolor{red}{+ 7.39 \%} & \textcolor{red}{+ 10.39 \%} & 0.0093 & \textcolor{red}{+ 6.02 \%} & \textcolor{red}{+ 9.09 \%} & 0.0118 & \textcolor{red}{+ 6.74 \%} & \textcolor{red}{+ 9.19 \%} \\
        & NDCG@10 & 0.0109 & \textcolor{red}{+ 6.01 \%} & \textcolor{red}{+ 8.13 \%} & 0.0113 & \textcolor{red}{+ 5.21  \%} & \textcolor{red}{+ 5.94 \%} & 0.0125 & \textcolor{red}{+ 4.32 \%} & \textcolor{red}{+ 8.07 \%} & 0.0155 & \textcolor{red}{+ 7.89 \%} & \textcolor{red}{+ 9.29 \%} \\
        
       \bottomrule
    \end{tabular}}
    
    \label{Main_results}
\end{table*}

\paragraph{Dataset.} To evaluate our proposed method, we conduct experiments on three benchmark datasets: the Amazon Beauty, Amazon-Clothing \cite{xue2022hypro,xue2023easytpp}, and MovieLens-1M (ML-1M) datasets. The Amazon datasets, originally introduced in \cite{mcauley2015image}, are known for high sparsity and short sequence lengths. We select the Beauty and Clothing subcategories, using the fine-grained product categories and brands as item attributes. The ML-1M dataset, from \cite{harper2015movielens}, is a large and dense
dataset consisting of long item sequences collected from the movie recommendation site MovieLens, with movie genres used as attributes. The statistics of the four datasets after preprocessing are summarized in Table \ref{Datasets}. Following common practice \cite{kang2018self,qiu2022contrastive,sun2019bert4rec,zhou2020s3}, we treat all user-item interactions as implicit feedback. For each user, we remove duplicate interactions and sort the remaining interactions chronologically to construct sequential user profiles. Through experiments on these diverse public datasets, we aim to thoroughly evaluate the performance of our proposed approach.

\paragraph{Evaluation Metrics.}
To evaluate the performance of our recommendation system, we utilize a leave-one-out strategy where we repeatedly hold out one item from each user's sequence of interactions. This allows us to test the model's ability to predict the held-out item. We make predictions over the entire item set without any negative sampling. We report two widely used ranking metrics - Top-$n$ metrics \textbf{HR@$n$} (Hit Rate) and \textbf{NDCG@$n$} (Normalized Discounted Cumulative Gain) where $n$ is set to 5 and 10. HR@$n$ measures whether the held-out item is present in the top-$n$ recommendations, while NDCG@$n$ considers the position of the held-out item by assigning higher scores to hits at top ranks. To ensure robust evaluation, we repeat each experiment 5 times with different random seeds and report the average performance across runs as the final metrics. This allows us to account for variability and ensure our results are not dependent on a particular random initialization.

\paragraph{Baselines.}
Following the experiment comparison \cite{du2023ensemble}, we include baseline methods from three groups for comparison: (1) General sequential methods utilize a sequence encoder to generate the hidden representations of users and items. For example, BERT4Rec \cite{sun2019bert4rec} adopts bidirectional Transformer as the sequence encoder. (2) Attribute-aware sequential methods fuse attribute information into sequential recommendation. For example, FDSA \cite{zhang2019feature} applies self-attention blocks to capture transition patterns of items and attribute. (3) Contrastive sequential methods design auxiliary objectives for contrastive learning based on general sequential methods. For example, CL4SRec \cite{xie2022contrastive} proposes data augmentation strategies for contrastive learning in sequential recommendation. DuoRec \cite{qiu2022contrastive} proposes both supervised and unsupervised sampling strategies for contrastive learning in sequential recommendation.

\begin{table*}[th!]
    \centering
    
    \caption{Ablation studies of our LLMRG model on two benchmark datasets, i.e., ML-1M and Amazon Beauty. We take the DuoRec as a baseline model to compare with the DuoRec with sequence graph and DuoRec with direct recommendation results via naive GPT3.5 or GPT4 without constructing a reasoning graph. The gray shaded area indicates our LLMRG's improved performance over the baseline across two datasets. The blue shaded area indicates the DuoRec with direct recommendation results via naive GPT3.5 or GPT4 without constructing a reasoning graph. Higher is better.}
    
    \resizebox{1.9\columnwidth}{!}{
        \begin{tabular}{lll |cccc |cccc }
        \toprule
       & \multirow{2.5}{*}{Method} & & \multicolumn{4}{c}{ML-1M} &
        \multicolumn{4}{c}{Amazon Beauty}  \\
        
          \cmidrule(lr){4-7} \cmidrule(lr){8-11} 
        
         &&&  HR@5& HR@10  & NDCG@5 & NDCG@10 &
           HR@5& HR@10  & NDCG@5 & NDCG@10\\
       
        \midrule
         DuoRec  & & & 0.2011 & 0.2837 & 0.1265 & 0.1663 & 0.0552 & 0.0839 & 0.0350 & 0.0447 \\
         
         DuoRec & w/ & seq graph & \textcolor{red}{+ 6.36 \%} & \textcolor{red}{+ 7.12 \%} & \textcolor{red}{+ 12.25 \%} & \textcolor{red}{+  4.50 \%} & \textcolor{red}{+ 3.26 \%} & \textcolor{red}{+ 2.74 \%} & \textcolor{red}{+ 3.71 \%} & \textcolor{red}{+ 2.68 \%} \\
         \rowcolor{blizzardblue}
         DuoRec+GPT3.5 & w/o & rea graph & \textcolor{red}{+ 0.94 \% }& \textcolor{red}{+ 0.81 \% } & \textcolor{red}{+ 0.55 \% } & \textcolor{red}{+  1.80 \% } & \textcolor{darkgray}{- 1.26 \%}& \textcolor{darkgray}{- 0.71 \%} & \textcolor{darkgray}{- 0.85 \%} & \textcolor{darkgray}{- 0.89 \%} \\
        \rowcolor{Gray}
        LLMRG (GPT3.5) & w/ & rea graph & \textcolor{red}{+ 12.87 \%} & \textcolor{red}{+ 14.10 \%} & \textcolor{red}{+ 23.55 \%} & \textcolor{red}{+  12.86 \%} & \textcolor{red}{+ 9.31 \%} & \textcolor{red}{+ 5.14 \%} & \textcolor{red}{+ 7.42 \%} & \textcolor{red}{+ 6.67  \%} \\
       \rowcolor{blizzardblue}
        DuoRec+GPT4 & w/o & rea graph & \textcolor{red}{+ 3.28  \%} & \textcolor{red}{+ 2.29 \%} & \textcolor{red}{+ 3.95 \%} & \textcolor{red}{+ 2.22 \%} & \textcolor{red}{+ 0.72 \%} & \textcolor{red}{+ 0.71 \%} & \textcolor{red}{+ 0.86 \%} & \textcolor{red}{+ 0.67 \%} \\
        \rowcolor{Gray}
        LLMRG (GPT4) & w/ & rea graph & \textcolor{red}{+ 14.76 \%} & \textcolor{red}{+ 15.53  \%} & \textcolor{red}{+ 26.01 \%} & \textcolor{red}{+ 13.77 \%} & \textcolor{red}{+ 11.93 \%} & \textcolor{red}{+ 6.61  \%} & \textcolor{red}{+ 9.24 \%} & \textcolor{red}{+ 7.95 \%} \\

       \bottomrule
    \end{tabular}}
    \label{results_graph}
\end{table*}

\begin{table*}
    \centering
    
    \caption{Ablation studies of our LLMRG model on two benchmark datasets, i.e., ML-1M and Amazon Beauty. We take the DuoRec as a baseline model to compare with the ablation models w/ or w/o divergent extension and self-verification modules based on GPT3.5 or GPT4. The shaded area indicates our ablation models' improved or decreased performance over the baseline across two datasets. Higher is better.}
    
    \resizebox{1.85\columnwidth}{!}{
        \begin{tabular}{l l |cccc |cccc }
        \toprule
       \multirow{2.5}{*}{LLM} & \multirow{2.5}{*}{Method} & \multicolumn{4}{c}{ML-1M} &
        \multicolumn{4}{c}{Amazon Beauty}  \\
        
          \cmidrule(lr){3-6} \cmidrule(lr){7-10} 
        
          & &  HR@5& HR@10  & NDCG@5 & NDCG@10 
         &  HR@5& HR@10  & NDCG@5 & NDCG@10\\
       
        \midrule
        NA & DuoRec & 0.2011 & 0.2837 & 0.1265 & 0.1663 & 0.0552 & 0.0839 & 0.0350 & 0.0447 \\
         \midrule
         \rowcolor{Gray}
         & w/o div & \textcolor{red}{+ 5.12 \%} & \textcolor{red}{+ 3.87 \%} & \textcolor{red}{+ 8.30 \%} & \textcolor{red}{+ 4.75 \%} & \textcolor{red}{+ 3.62 \%} & \textcolor{red}{+ 2.86 \%} & \textcolor{red}{+ 4.57 \%} & \textcolor{red}{+ 3.80 \%} \\
         \rowcolor{Gray}
         GPT3.5 & w/o ver & \textcolor{darkgray}{- 4.72 \%} & \textcolor{darkgray}{- 3.94 \%} & \textcolor{darkgray}{- 10.90 \%} & \textcolor{darkgray}{- 4.14 \%} & \textcolor{darkgray}{- 2.17 \%} & \textcolor{darkgray}{- 1.43 \%} & \textcolor{darkgray}{- 2.57 \%} & \textcolor{darkgray}{- 2.46 \%} \\
        \rowcolor{Gray}
        & w/ div \& ver & \textcolor{red}{+ 12.87 \%} & \textcolor{red}{+ 14.10 \%} & \textcolor{red}{+ 23.55 \%} & \textcolor{red}{+ 12.86 \%} & \textcolor{red}{+ 9.31 \%} & \textcolor{red}{+ 5.14 \%} & \textcolor{red}{+ 7.42 \%} & \textcolor{red}{+ 6.67 \%} \\
       \midrule
       \rowcolor{Gray}
        & w/o div & \textcolor{red}{+ 7.06 \%} & \textcolor{red}{+ 4.68 \%} & \textcolor{red}{+ 13.35 \%} & \textcolor{red}{+ 8.11 \%} & \textcolor{red}{+ 4.89 \%} & \textcolor{red}{+ 3.45 \%} & \textcolor{red}{+ 4.28 \%} & \textcolor{red}{+ 5.81 \%} \\
        \rowcolor{Gray}
        GPT4 & w/o ver & \textcolor{red}{+ 5.86 \%} & \textcolor{red}{+ 2.36 \%} & \textcolor{red}{+ 5.77  \%} & \textcolor{red}{+ 3.72 \%} & \textcolor{red}{+ 1.26 \%} & \textcolor{red}{+ 1.31 \%} & \textcolor{red}{+ 1.71 \%} & \textcolor{red}{+ 1.56 \%} \\
        \rowcolor{Gray} 
        & w/ div \& ver & \textcolor{red}{+ 14.76 \%} & \textcolor{red}{+ 15.53 \%} & \textcolor{red}{+ 26.01 \%} & \textcolor{red}{+ 13.77 \%} & \textcolor{red}{+ 11.93 \%} & \textcolor{red}{+ 6.61 \%} & \textcolor{red}{+ 9.24 \%} & \textcolor{red}{+ 7.95 \%} \\

       \bottomrule
    \end{tabular}}
    
    \label{results_div_ver}
\end{table*}

\begin{figure*}[t]
    \centering
    \includegraphics[width=2\columnwidth]{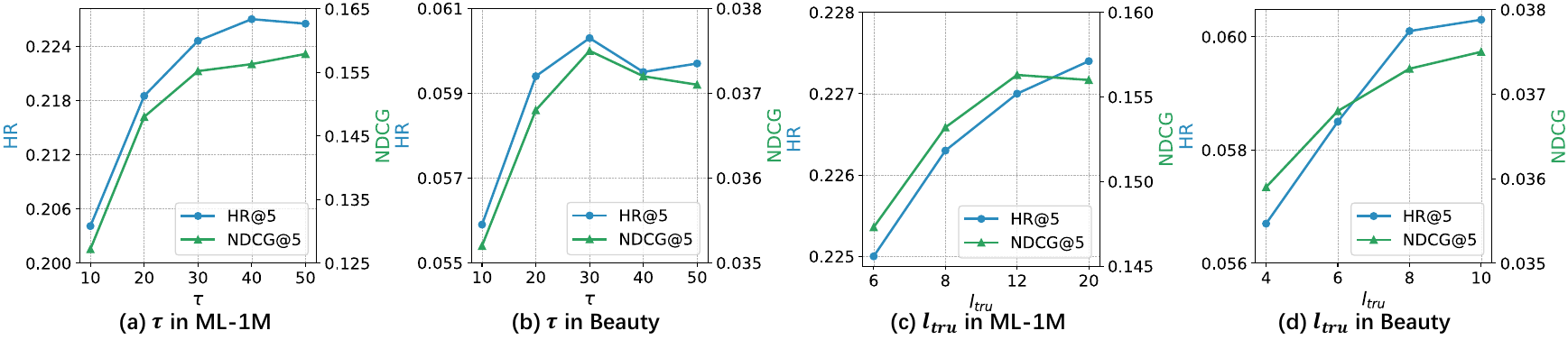}
    
    \caption{Sensitivity analysis of threshold of verification scoring $\tau$ and sequence truncation length $l_{tru}$ on HR and NDCG performance based on ML-1M and Amazon Beauty benchmarks.}
    
    \label{fig:exp_tau_length}
\end{figure*}

\begin{figure}[t]
    \centering
    \includegraphics[width=0.7\columnwidth]{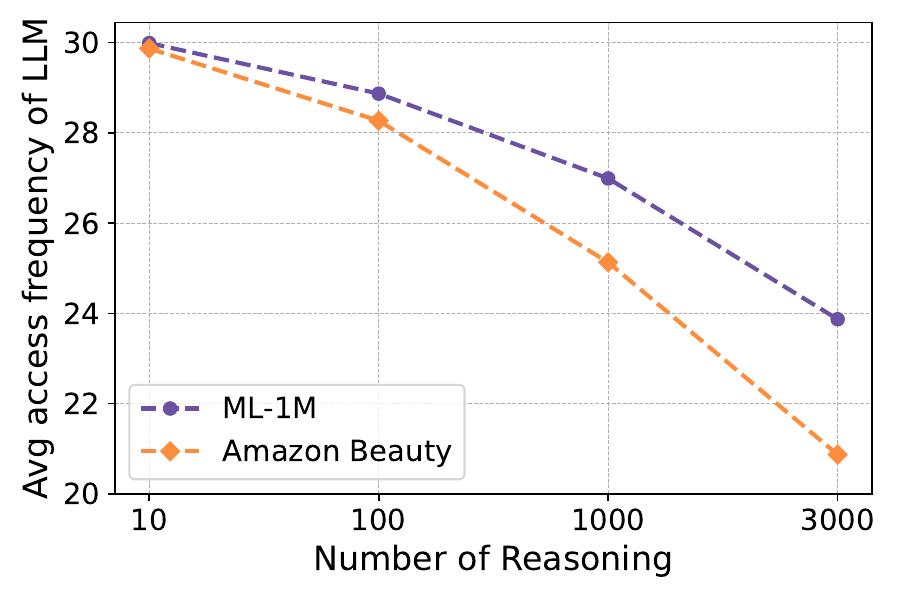}
    
    \caption{The average access frequency of LLM  based on ML-1M and Amazon Beauty benchmarks.}
    
    \label{fig:exp_self_improving}
\end{figure}

\paragraph{Results and Analysis.}
As evidenced in Table \ref{Main_results}, we conduct comprehensive benchmarking experiments on three widely-used datasets - ML-1M, Amazon Beauty, and Amazon Clothing. The detailed experiment results with absolute increase values are provided in the Appendix. We compare our proposed LLMRG model built on top of GPT3.5 or GPT4 with several strong baseline methods, including FDSA \cite{zhang2019feature}, BERT4Rec \cite{sun2019bert4rec}, CL4SREC \cite{xie2022contrastive}, and DuoRec \cite{qiu2022contrastive}. The shaded regions in the table highlight the performance improvements achieved by our LLMRG model over all baselines across the three datasets. These results demonstrate the plug-and-play nature of LLMRG, which can effectively enhance multiple existing recommenders. More importantly, we observe significant performance gains on HR@5, HR@10, NDCG@5, and NDCG@10 after applying LLMRG, compared to the original baseline models. This indicates that conventional recommender systems struggle to model the conceptual relationships and behavioral sequences of diverse user interests. In contrast, our proposed LLMRG framework can boost recommendation performance without needing any additional information. These improvements showcase how large language models can bring logical reasoning and interpretability to recommender systems. Furthermore, LLMRG performance scales with the underlying LLM capability - the GPT4-based LLMRG consistently outperforms its GPT3.5 counterpart. In addition, when comparing the ML-1M movie dataset to the Beauty and Clothing product datasets, we observed that our LLMRG approach led to greater improvements across all evaluation metrics on the ML-1M dataset. This suggests that movie items contain richer semantic information and enable more semantically logical reasoning relationships than Amazon product items. As movies often have complex plots, character arcs, and artistic themes, recommending movies likely requires more sophisticated relational reasoning between items than recommending simple retail products. The complexity of logical relations between movie entities enables our LLMRG method to better leverage its relational modeling capabilities. In contrast, beauty and clothing products have less narrative complexity, so there is less opportunity for relational reasoning to improve recommendations.

\paragraph{Ablation Study.}

To demonstrate the effectiveness of our proposed reasoning graph, we conduct ablation studies on our LLMRG model using two benchmark datasets: ML-1M and Amazon Beauty. We compare LLMRG to the DuoRec baseline model as well as DuoRec augmented with a simple sequence graph, as proposed by \citet{wu2019session}. The sequence graph directly models interaction sequences without reasoning. We also compare against combining DuoRec with large language models - GPT-3.5 and GPT-4 - without constructing a reasoning graph. Here, the LLM simply outputs recommended items based on prompts containing historical sequences and user profiles, without reasoning graph. As shown in Figure \ref{results_graph}, the DuoRec model augmented with a sequence graph provides only minor improvements compared to our full LLMRG model. The DuoRec+GPT3.5 model without reasoning graph integration fails to significantly improve DuoRec performance on ML-1M, and even decreases performance on the Amazon Beauty dataset. Thanks to its greater capability, DuoRec+GPT4 boosts performance over DuoRec+GPT3.5 but still lags far behind our LLMRG model. These results demonstrate that the reasoning graph constructed by our proposed instructions is critical for performance, and simple next-item prediction is insufficient (DuoRec+GPT3.5 and DuoRec+GPT4). By explicitly modeling the reasoning process between user profiles and interaction sequences, LLMRG is able to make accurate, explainable recommendations. Our ablation studies confirm the reasoning graph's necessity and value in effectively leveraging the power of large language models for recommendation systems.

Our additional ablation studies further explore the effectiveness of each module in our LLMRG framework. Using DuoRec as a baseline model, we compared it to ablation versions of LLMRG with or without the divergent extension and self-verification modules based on GPT3.5 or GPT4. The results in Table \ref{results_div_ver} reveal that LLMRG (with GPT3.5 or GPT4) without the divergent extension module provides only marginal improvement compared with the complete LLMRG. However, removing the self-verification module from LLMRG (GPT3.5) actually decreases performance. This demonstrates the limited reasoning capability of GPT3.5 - without verification, uncontrolled reasoning introduces noise that reduces overall performance. Overall, these ablation experiments clearly demonstrate the value of both our divergent extension and self-verification modules in enabling more advanced reasoning while maintaining accuracy. The modules work synergistically to expand the search space of possible solutions while filtering out inaccurate or incoherent lines of reasoning.

As shown in Figure \ref{fig:exp_self_improving}, we also analyze the effectiveness of the proposed knowledge base self-improving. Based on LLMRG (GPT3.5), we calculate the average access frequency of the model call to LLM on two benchmark datasets. The experimental results show that the average access frequency decreases significantly as the number of reasoning steps increases. After 3,000 times of reasoning and verification, the average access frequency decreases by about $30\%$ compared to not using this module, proving that the knowledge base contains high-quality reasoning chains that can be reused. Moreover, we observed that the reuse rate of high-quality reasoning chains in Amazon Beauty is higher than that of ML-1M, and the long-tailed distribution of Amazon products is one of the reasons for this difference.

To provide intuitive examples corroborating our quantitative results, we examined real case studies from the ML-1M dataset using (a) our complete LLMRG model, (b) LLMRG without the divergent extension module, and (c) LLMRG without the self-verification module. The case studies in Figure \ref{fig:real_case_example} and Appendix illustrate the differences in reasoning between the models. LLMRG generates coherent recommendations with sound justifications, leveraging both divergent thinking to expand possibilities and self-verification to filter out poor options. Without divergent extensions, LLMRG struggles to move beyond obvious choices. And without self-verification, LLMRG's recommendations become more speculative and sometimes nonsensical, as the model lacks the ability to check its own thinking. These qualitative analyses mirror the patterns in our numerical results, serving as further validation of the value added by each reasoning module working in concert with our full LLMRG framework. The case studies provide intuitive examples of how our approach combines creative thinking and critical evaluation to produce logical recommendations.

\paragraph{Sensitivity Analysis}
We evaluate LLMRG's sensitivity to the two most crucial parameters, $\tau$ and $l_{tru}$, on HR and NDCG, which control the threshold for verification scoring and sequence truncation length, respectively. Figure~\ref{fig:exp_tau_length} (a) and (b) show that larger $\tau$ values yield more robust reasoning and filter out inferior options, thus boosting the model's performance on the ML-1M dataset. However, on the Beauty dataset, performance starts to decrease from $\tau=30$, likely because higher verification scoring thresholds filter out more reasoning chains, increasing the sparsity of the graph. Figures~\ref{fig:exp_tau_length} (c) and (d) indicate that, generally, longer sequences bring better recommendation results by incorporating more information. In summary, larger $\tau$ and longer sequences both tend to improve performance. $\tau$ exhibits a peak value, beyond which sparser reasoning graphs degrade results, especially for less logical sequences, such as Amazon products.

\section{Conclusion}
In this paper, we present LLMRG, a novel approach that utilizes LLM to construct personalized reasoning graphs. This method demonstrates how LLM can bring logical reasoning and interpretability to recommendation systems without needing any additional information.  We verify the effectiveness of our LLMRG method using real-world datasets and demonstrate that our plug-and-play method can effectively enhance multiple existing recommenders. However, it should be pointed out that although knowledge base self-improving is designed, LLMRM is essentially limited by the frequency of LLM access for longer interaction sequences.

\bibliography{aaai24}

\clearpage
\appendix

\begin{table*}[t]
    \centering
    \caption{Performance comparison on three benchmark datasets, i.e., ML-1M, Amazon Beauty, and Amazon Clothing. We set the original models as baselines to compare with our proposed LLMRG model based on GPT3.5 or GPT4. The shaded area indicates the improved performance of our LLMRG model over the baselines across all three datasets. Higher is better.}
    \resizebox{1.9\columnwidth}{!}{
        \begin{tabular}{l l|baa |baa |baa |baa}
        \toprule
        \multirow{2.5}{*}{Dataset} & \multirow{2.5}{*}{Metric} &
        \multicolumn{3}{c}{FDSA} & \multicolumn{3}{c}{BERT4Rec} & 
        \multicolumn{3}{c}{CL4SRec} & \multicolumn{3}{c}{DuoRec} \\
        
          \cmidrule(lr){3-5} \cmidrule(lr){6-8} \cmidrule(lr){9-11} \cmidrule(lr){12-14}
        
          &  & Original  & GPT3.5 & GPT4 
         & Original  & GPT3.5 & GPT4
         & Original  & GPT3.5 & GPT4
         & Original  & GPT3.5 & GPT4\\
       
        \midrule
         \multirow{4}{4em}{ML-1M} & HR@5 & 0.0909  & 0.1097  & 0.1143 & 0.1124 & 0.1423 & 0.1490 & 0.1141 & 0.1369 & 0.1380 & 0.2011 & 0.2270 & 0.2307 \\
         & HR@10 & 0.1631 &  0.1923 & 0.2004 & 0.1910 & 0.2168 & 0.2225 & 0.1866 & 0.2189 & 0.2226 & 0.2837 & 0.3237 & 0.3277 \\
        & NDCG@5 & 0.0599 & 0.0726 & 0.0780 & 0.0713 &  0.0896 & 0.0947 & 0.0721 &  0.0829 & 0.0841 & 0.1265 & 0.1563 & 0.1594 \\
        & NDCG@10 & 0.0878 & 0.1069 & 0.1126 & 0.0980 &  0.1208 &  0.1255 & 0.1013 &  0.1192 & 0.1219 & 0.1663 & 0.1877 & 0.1891 \\
         \midrule
        & HR@5 & 0.0237 & 0.0269& 0.0278& 0.0201 & 0.0239 & 0.0247 & 0.0398 & 0.0442 & 0.0454 & 0.0552 & 0.0603 &  0.0617 \\
        Amazon & HR@10 & 0.0418  & 0.0480 & 0.0492 & 0.0413 & 0.0486 & 0.0504 & 0.0664 & 0.0731 & 0.0739 & 0.0839 & 0.0882  & 0.0894 \\
        Beauty & NDCG@5 & 0.0195 & 0.0226 & 0.0231 & 0.0192 & 0.0219 &0.0225 & 0.0221 & 0.0239 & 0.0243 & 0.0350 & 0.0375 & 0.0382 \\
        & NDCG@10 & 0.0275 & 0.0315 & 0.0323 & 0.0263 & 0.0293 & 0.0301 & 0.0322 & 0.0348 & 0.0353 & 0.0447 & 0.0476 &  0.0482\\
        \midrule
        & HR@5 &  0.0119 & 0.0143 & 0.0147 & 0.0128 & 0.0148 & 0.0152 & 0.0166 & 0.0179 & 0.0184 & 0.0190 & 0.0208 &  0.0211 \\
       Amazon  & HR@10 & 0.0197 & 0.0225 & 0.0232 & 0.0202 & 0.0223 & 0.0229 & 0.0273 & 0.0303 & 0.0313 & 0.0311 &  0.0334 &  0.0340  \\
    Clothing & NDCG@5 & 0.0073 & 0.0078 & 0.0080 & 0.0081 &  0.0086 &  0.0089 & 0.0093 & 0.0098 & 0.0101 & 0.0118 & 0.0125 & 0.0128 \\
        & NDCG@10 & 0.0109 &  0.0115 & 0.0117 & 0.0113 & 0.0118 & 0.0119 & 0.0125 &  0.0130 & 0.0135  & 0.0155 &  0.0167 & 0.0169  \\
        
       \bottomrule
    \end{tabular}}
    \label{Main_results_value}
\end{table*}

\begin{table*}[t]
    \centering
    \caption{Ablation studies of our LLMRG model on two benchmark datasets, i.e., ML-1M and Amazon Beauty. We take the DuoRec as a baseline model to compare with the ablation models w/ or w/o divergent extension and self-verification modules based on GPT3.5 or GPT4. The shaded area indicates our ablation models' improved or decreased performance over the baseline across two datasets. Higher is better.}
    \resizebox{1.9\columnwidth}{!}{
        \begin{tabular}{l l |cccc |cccc }
        \toprule
       \multirow{2.5}{*}{LLM} & \multirow{2.5}{*}{Method} & \multicolumn{4}{c}{ML-1M} &
        \multicolumn{4}{c}{Amazon Beauty}  \\
        
          \cmidrule(lr){3-6} \cmidrule(lr){7-10} 
        
          & &  HR@5& HR@10  & NDCG@5 & NDCG@10 
         &  HR@5& HR@10  & NDCG@5 & NDCG@10\\
       
        \midrule
        NA & DuoRec & 0.2011 & 0.2837 & 0.1265 & 0.1663 & 0.0552 & 0.0839 & 0.0350 & 0.0447 \\
         \midrule
         \rowcolor{Gray}
         & w/o div & 0.2114 & 0.2947 & 0.1370 &0.1742  & 0.0572 & 0.0863 &  0.0366 & 0.0464 \\
         \rowcolor{Gray}
         GPT3.5 & w/o ver & 0.1916 & 0.2725 & 0.1127 & 0.1594  & 0.0540 & 0.0827  & 0.0341 &  0.0436 \\
        \rowcolor{Gray}
        & w/ div \& ver & 0.2270 & 0.3237  & 0.1563 &0.1877  & 0.0603 & 0.0882 & 0.0375 &  0.0476 \\
       \midrule
       \rowcolor{Gray}
        & w/o div & 0.2153 & 0.2970 &  0.1434&  0.1798& 0.0579 & 0.0868  & 0.0365 & 0.0473\\
        \rowcolor{Gray}
        GPT4 & w/o ver &  0.2129 & 0.2904 & 0.1338 &0.1725 & 0.0559 & 0.0850 & 0.0356 & 0.0454 \\
        \rowcolor{Gray} 
        & w/ div \& ver &0.2307 & 0.3277  & 0.1594 & 0.1891 &0.0617 & 0.0894 & 0.0382 &  0.0482  \\

       \bottomrule
    \end{tabular}}
    \label{results_div_ver_value}
\end{table*}

\begin{table*}[t]
    \centering
    \caption{Ablation studies of our LLMRG model on two benchmark datasets, i.e., ML-1M and Amazon Beauty. We take the DuoRec as a baseline model to compare with the DuoRec with sequence graph and DuoRec with direct recommendation results via naive GPT3.5 or GPT4 without constructing a reasoning graph. The gray shaded area indicates our LLMRG's improved performance over the baseline across two datasets. The blue shaded area indicates the DuoRec with direct recommendation results via naive GPT3.5 or GPT4 without constructing a reasoning graph. Higher is better.}
    \resizebox{1.9\columnwidth}{!}{
        \begin{tabular}{lll |cccc |cccc }
        \toprule
       & \multirow{2.5}{*}{Method} & & \multicolumn{4}{c}{ML-1M} &
        \multicolumn{4}{c}{Amazon Beauty}  \\
        
          \cmidrule(lr){4-7} \cmidrule(lr){8-11} 
        
         &&&  HR@5& HR@10  & NDCG@5 & NDCG@10 &
           HR@5& HR@10  & NDCG@5 & NDCG@10\\
       
        \midrule
         DuoRec  & & & 0.2011 & 0.2837 & 0.1265 & 0.1663 & 0.0552 & 0.0839 & 0.0350 & 0.0447 \\
         
         DuoRec & w/ & seq graph & 0.2139 & 0.3039 & 0.1420 & 0.1738 & 0.0570 & 0.0862  & 0.0363  &  0.0459 \\
         \rowcolor{blizzardblue}
         GPT3.5(naive) & w/o & rea graph & 0.2030 & 0.2860  &0.1272 & 0.1693 & 0.0545 & 0.0833 & 0.0347 & 0.0443\\
        \rowcolor{Gray}
        GPT3.5(LLMRG) & w/ & rea graph & 0.2270 & 0.3237 &  0.1563 &  0.1877 &  0.0603 & 0.0882 & 0.0375 & 0.0476 \\
       \rowcolor{blizzardblue}
        GPT4(naive) & w/o & rea graph &  0.2077 & 0.2902 & 0.1315 & 0.1700 &  0.0556 &  0.0845 &  0.0353 & 0.0450 \\
        \rowcolor{Gray}
        GPT4(LLMRG) & w/ & rea graph & 0.2307  & 0.3277 & 0.1594 & 0.1891 & 0.0617 & 0.0894 & 0.0382 & 0.0482 \\

       \bottomrule
    \end{tabular}}
    \label{results_graph_value}
\end{table*}

\begin{figure*}[t]
    \centering
    \includegraphics[width=1.8\columnwidth]{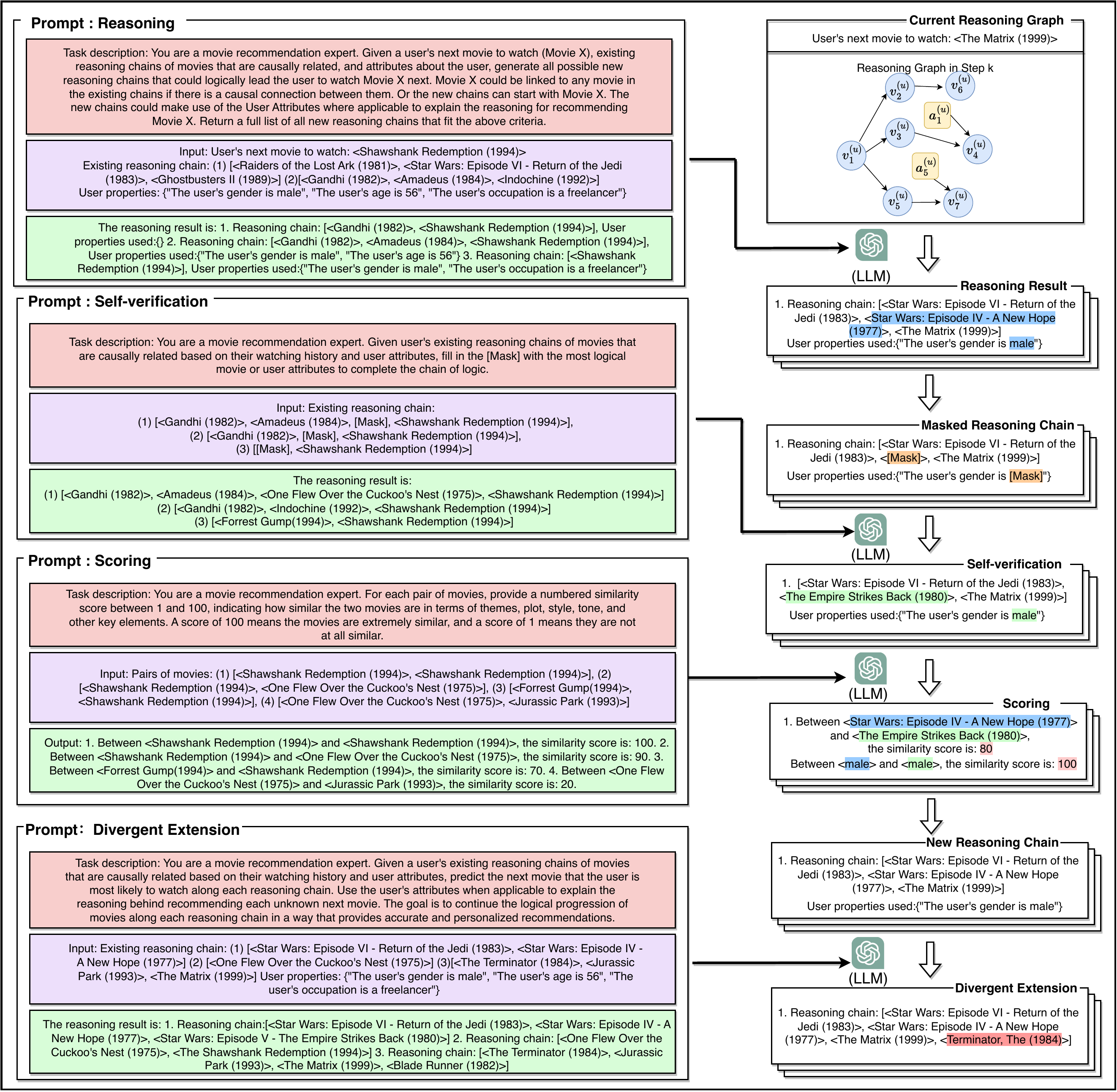}
    \caption{The prompt examples and real cases for graph reasoning, self-verification, scoring, and divergent extension modules. This is the input structure of LLM, which is accessible via a string to the API and divided into three components: task description, example input, and example output. The task description appears first and indicates the type of task being requested, providing critical context that constrains LLM's behavior. The example input demonstrates specific content that LLM should respond to this task. Finally, the example output illustrates the desired form the reply should take for this prompt.}
    \label{fig:prompt_example}
\end{figure*}

\begin{figure*}[t]
    \centering
    \includegraphics[width=2\columnwidth]{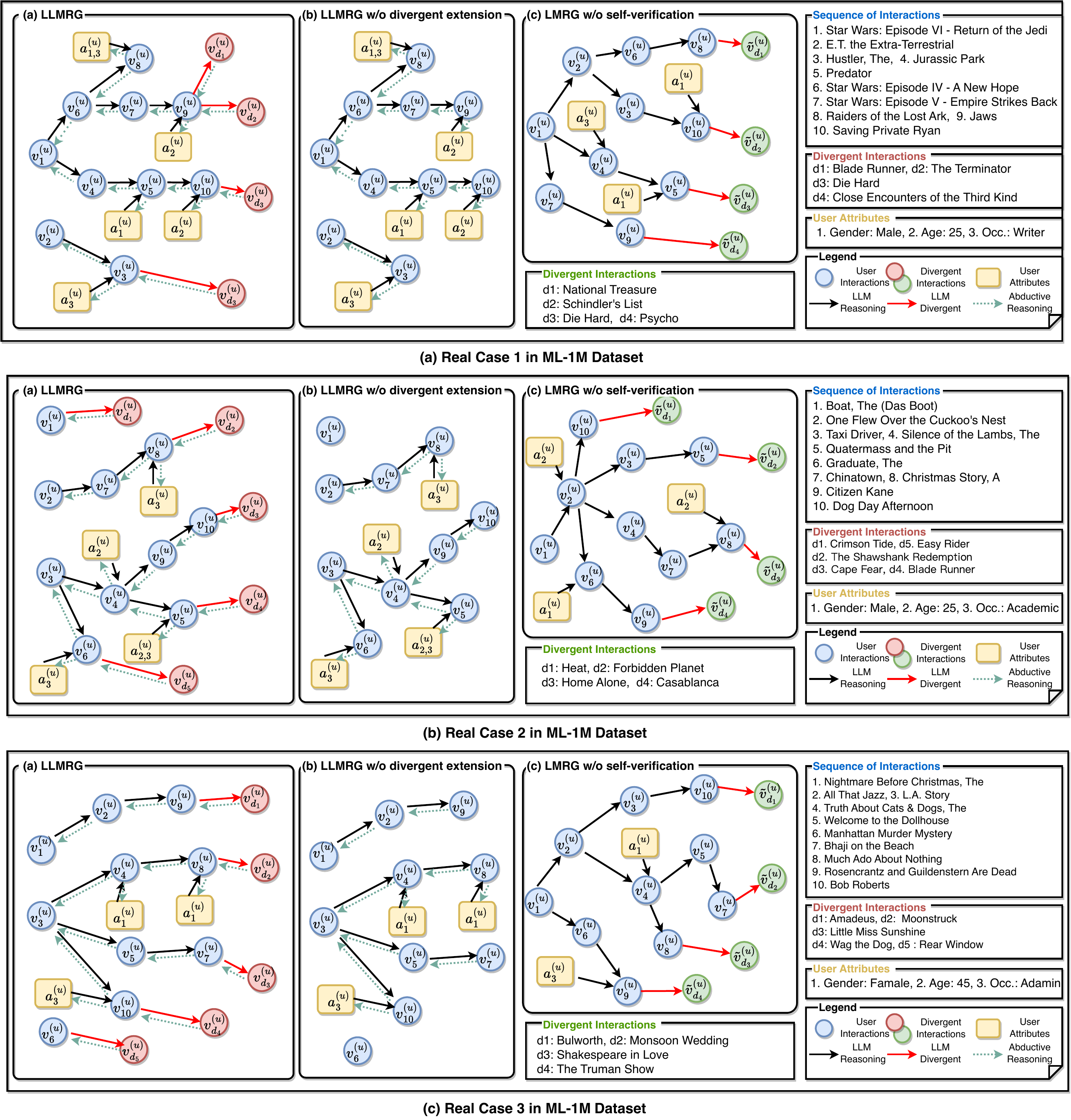}

    \caption{The real case studies (ML-1M) on our (a) LLMRG and ablation models, i.e., (b) LLMRG w/o divergent extension and (c) LLMRG w/o self-verification. The black arrow represents the reasoning procedure. The red arrow is the divergent extension. The green dashed arrow refers to the abductive reasoning in the self-verification module.}

    \label{fig:app_real_case_all}
\end{figure*}

\end{document}